\documentclass{elsart}
\usepackage{amssymb}
\usepackage{graphicx}

\begin{document}

\newcommand\cc{{\rm CC}}
\newcommand\nc{{\rm NC}}
\newcommand\nutau{{\nu_\tau}}
\newcommand\anutau{\bar{\nu}_\tau}
\newcommand\numu{{\nu_\mu}}
\newcommand\anumu{\bar{\nu}_\mu}
\newcommand\nue{{\nu_e}}
\newcommand\anue{\bar{\nu}_e}

\newcommand\enunu{e^- \anue \nutau}
\newcommand\mununu{\mu^- \anumu \nutau}
\newcommand\taue{\tau^- \to \enunu}
\newcommand\taumu{\tau^- \to \mununu}

\newcommand{\ra}{\rightarrow}
\newcommand{\nmne}{\numu \ra \nue}
\newcommand{\anmane}{\anumu \ra \anue}
\newcommand{\dm}{\Delta m^2}
\newcommand{\sintot}{\sin^2(2\theta)}
\newcommand{\Rem}{R_{e \mu}}
\newcommand{\Eh}{E_h}
\newcommand{\qlep}{Q_{\rm lep}}
\newcommand{\phieh}{\phi_{\rm eh}}
\newcommand{\Ybj}{Y_{\rm Bj}}
\newcommand{\Ebrem}{E_{\rm brem}}

\newcommand{\pip}{\pi^+}
\newcommand{\pim}{\pi^-}
\newcommand{\piz}{\pi^0}
\newcommand{\pis}{\pi^\pm}
\newcommand{\kp}{K^+}
\newcommand{\km}{K^-}
\newcommand{\ks}{K^\pm}
\newcommand{\klong}{K^0_L}
\newcommand{\kshort}{K^0_S}
\newcommand{\mup}{\mu^+}
\newcommand{\mum}{\mu^-}
\newcommand{\mus}{\mu^\pm}

\newcommand{\epsdir}{.}

\newcommand{\fix}[1]{{\bf <<< #1 !!! }}

\begin{frontmatter}

\title{\boldmath Search for $\nmne$ oscillations in the NOMAD experiment}
%\collab{\bf NOMAD Collaboration} % Do not use - conflicts with \corauth
{\large NOMAD Collaboration}

\author[Paris]             {P.~Astier},
\author[CERN]              {D.~Autiero},
\author[Saclay]            {A.~Baldisseri},
\author[Padova]            {M.~Baldo-Ceolin},
\author[Paris]             {M.~Banner},
\author[LAPP]              {G.~Bassompierre},
\author[Lausanne]          {K.~Benslama},
\author[Saclay]            {N.~Besson},
\author[CERN,Lausanne]     {I.~Bird},
\author[Johns Hopkins]     {B.~Blumenfeld},
% ...
\corauth[corr]                {Corresponding author.}
\ead                          {Slava.Valouev@cern.ch}
\author[Paris]             {F.~Vannucci},
\author[Sydney]            {K.E.~Varvell},
\author[Urbino]            {M.~Veltri},
\author[Pavia]             {V.~Vercesi},
\author[CERN]              {G.~Vidal-Sitjes},
\author[Lausanne]          {J.-M.~Vieira},
\author[UCLA]              {T.~Vinogradova},
\author[Harvard,CERN]      {F.V.~Weber},
\author[Dortmund]          {T.~Weisse},
\author[CERN]              {F.F.~Wilson},
\author[Melbourne]         {L.J.~Winton},
\author[Sydney]            {B.D.~Yabsley},
\author[Saclay]            {H.~Zaccone},
\author[Dortmund]          {K.~Zuber},
\author[Padova]            {P.~Zuccon}

\address[LAPP]           {LAPP, Annecy, France}
\address[Johns Hopkins]  {Johns Hopkins Univ., Baltimore, MD, USA}
\address[Harvard]        {Harvard Univ., Cambridge, MA, USA}
% ...
\address[Dortmund]       {Dortmund Univ., Dortmund, Germany}
% ...
\address[CERN]           {CERN, Geneva, Switzerland}
\address[Lausanne]       {University of Lausanne, Lausanne, Switzerland}
\address[UCLA]           {UCLA, Los Angeles, CA, USA}
\address[Melbourne]      {University of Melbourne, Melbourne, Australia}
% ...
\address[Padova]         {Univ. of Padova and INFN, Padova, Italy}
\address[Paris]          {LPNHE, Univ. of Paris VI and VII, Paris, France}
\address[Pavia]          {Univ. of Pavia and INFN, Pavia, Italy}
% ...
\address[Saclay]         {DAPNIA, CEA Saclay, France}
% ...
\address[Sydney]         {Univ. of Sydney, Sydney, Australia}
\address[Urbino]         {Univ. of Urbino, Urbino, and INFN Florence, Italy}
% ...

%\clearpage
\begin{abstract}
  We present the results of a search for $\nmne$ oscillations
  in the NOMAD experiment at CERN.  The experiment looked for the
  appearance of $\nue$ in a predominantly $\numu$ wide-band neutrino
  beam at the CERN SPS.  No evidence for oscillations was found.
  The 90\% confidence limits obtained are $\dm < 0.4$~eV$^2$
  for maximal mixing and $\sintot < 1.4 \times 10^{-3}$ for large $\dm$.
  This result excludes the LSND allowed region of oscillation parameters
  with $\dm \gtrsim 10$~eV$^2$.
\end{abstract}


\end{frontmatter}

\section{Introduction}
The NOMAD experiment was designed to search for $\nutau$ appearance from
neutrino oscillations in the CERN wide-band neutrino beam produced by the
450~GeV proton synchrotron (SPS).  The detector was optimized to identify
efficiently electrons from $\taue$ decays and
therefore could also be used to look for $\nue$ appearance in a
predominantly $\numu$ beam by detecting their charged current (\cc)
interactions $\nue N \ra e^- X$.  The main motivation for this search
was the evidence for $\anmane$ and $\nmne$ oscillations found by
the LSND experiment~\cite{LSND}.  For $\nmne$ oscillations with
$\dm \gtrsim 10$~eV$^2$ and with the probability of $2.6 \times 10^{-3}$
observed by LSND, a signal should be seen in the NOMAD data.  The sensitivity
of the NOMAD experiment to lower values of $\dm$ is limited by its $L/E_{\nu}$
ratio of $\sim$ 0.025~km/GeV, where $L$ is the average source to detector
distance and $E_{\nu}$ is the average neutrino energy.

Preliminary results of the search for $\nmne$ oscillations in NOMAD were
presented earlier~\cite{EPS01}.  In this paper we report the final results
of our ``blind'' analysis. 

\section{NOMAD detector and data collection}
The NOMAD detector~\cite{NOMAD} consisted of a large dipole magnet
delivering a field of 0.4~T and housing several subdetectors, starting
with an active target composed of 132 planes of drift chambers (DC) of
3$\times$3~m$^2$~\cite{DC}.  The walls of the chambers provided a low average
density (0.1 g/cm$^3$) target with a mass of 2.7 tons.  The density of the
chambers was low enough to allow an accurate measurement of the momenta
of the charged particles produced in the neutrino interactions.  The
chambers were followed by nine transition radiation (TRD) modules~\cite{TRD}
each consisting of a polypropylene radiator and a plane of straw
tubes operated with an 80\% xenon and 20\% methane gas mixture.
An electromagnetic calorimeter (ECAL) consisting of 875 lead glass
blocks~\cite{ECAL,ECAL_nonlin} provided a measurement of the energies
of electrons and
photons with a resolution of $\sigma(E)/E=3.2\%/\sqrt{E(\mbox{GeV})} + 1\%$.
The ECAL was preceded by a lead-proportional tube preshower for better
photon localization.  A hadron calorimeter (HCAL) was located just
beyond the magnet coil and was followed by two muon stations consisting
of large area drift chambers, the first station located after 8, and
the second one after 13 interaction lengths of iron.  Two planes
of scintillator counters, $T_1$ and $T_2$, were placed before and after
the TRD modules.  A third scintillator plane, $V$, placed upstream
of the magnet, was used to veto interactions caused by incoming
charged particles.  The trigger~\cite{TRIG} used in this analysis was
$\bar{V} \times T_1 \times T_2$.

\section{Neutrino beam}\label{sec:beam}
The CERN West Area Neutrino Facility (WANF) neutrino beam~\cite{WANF}
was produced
by impinging 450 GeV protons extracted from the SPS onto a target
consisting of beryllium rods adding up to a total thickness of 110~cm.
The secondary particles emerging
from the target were focused into a near parallel beam by two magnetic
lenses (the horn and the reflector) providing toroidal magnetic fields.
When running in neutrino mode positively charged particles were focused.
When running in the antineutrino mode the polarity of the two lenses
was reversed thus focusing negatively charged
particles.  The focused particles then traversed a 290~m long decay
tunnel followed by an iron and earth shield.  Neutrinos originating
from the decay of these particles travelled on average a distance
of 625~m before reaching the NOMAD detector.

Since the oscillation search implies a direct comparison between the
measured and expected ratios of the number of $\nue$~CC to $\numu$~CC
interactions, an accurate prediction of the neutrino
fluxes and spectra is crucial.  They are computed with a detailed Monte
Carlo simulation of the neutrino beam, referred to as NUBEAM and described
in Ref.~\cite{fluxes}.  This is implemented
in three steps.  First, the yields of the secondary particles from the
interactions of 450 GeV protons with the Be target are calculated
with the 2000 version of FLUKA~\cite{FLUKA}, a generator of hadronic
interactions. These yields are then modified in order to agree with
all measurements
currently available in the relevant energy and angular range, namely
the SPY/NA56~\cite{SPY} and NA20~\cite{Atherton} results.  Finally,
the propagation of the secondary particles
is described by a simulation program based on GEANT3~\cite{GEANT}, which
includes
an accurate description of the magnetic field in the horn and reflector,
and the modelling of reinteractions in the beam elements.

\begin{figure}
  \begin{center}
    \begin{minipage}[c]{0.5\textwidth}
     \centering\includegraphics[width=1.\textwidth]{\epsdir/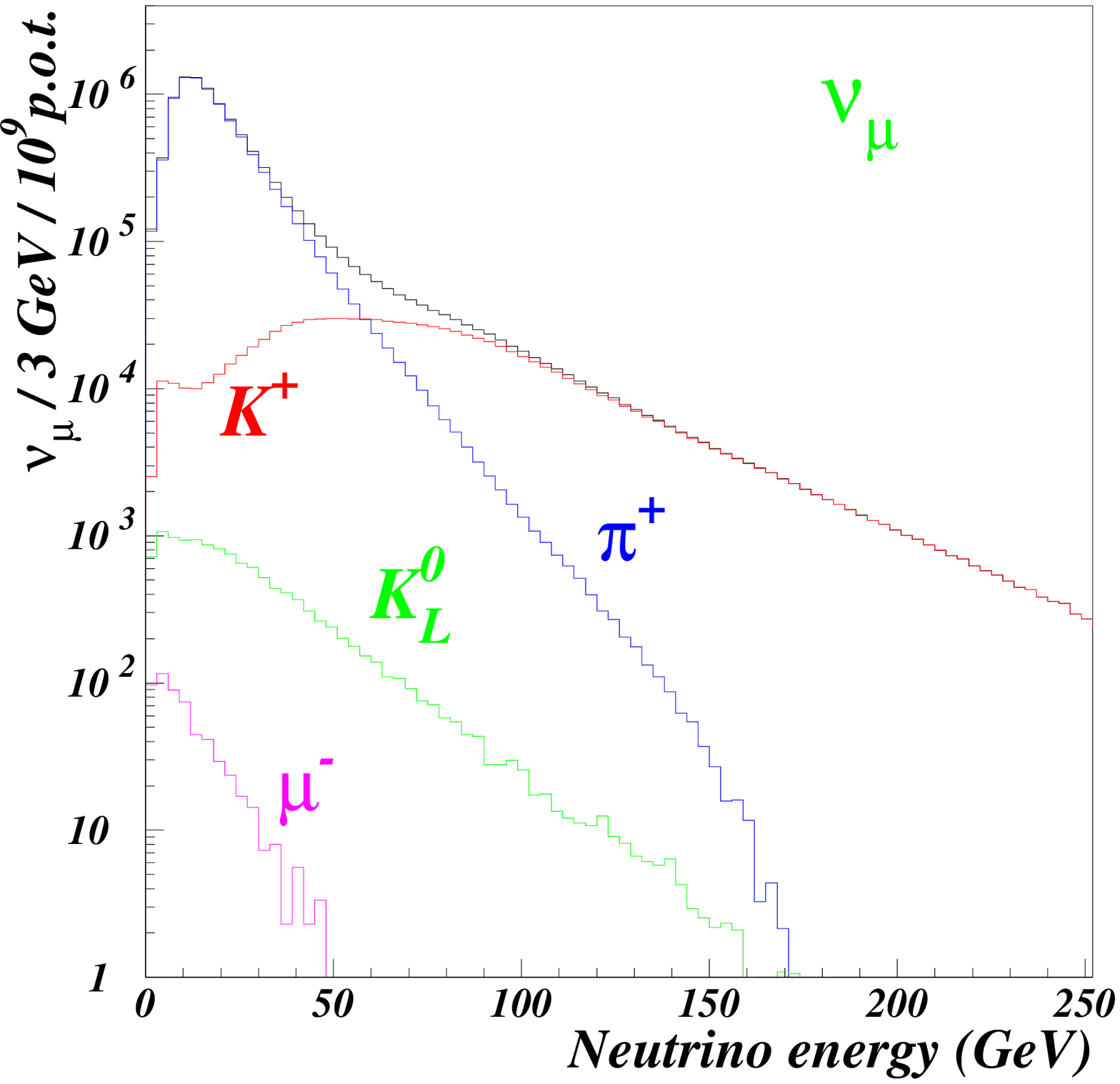}
    \end{minipage}\hfill
    \begin{minipage}[c]{0.5\textwidth}
     \centering\includegraphics[width=1.\textwidth]{\epsdir/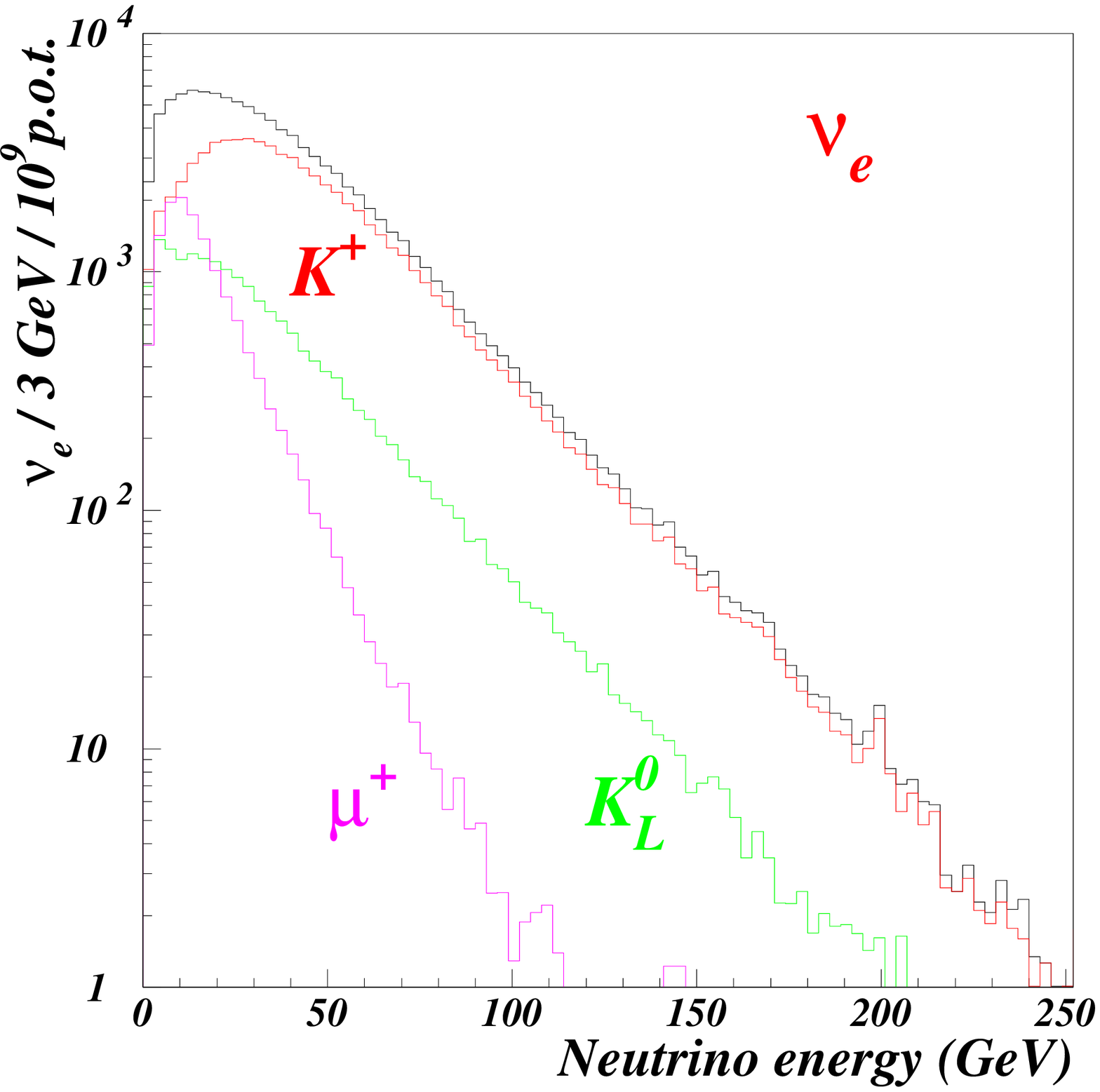}
    \end{minipage}
  \end{center}
  \caption{Composition of the $\numu$ and $\nue$ energy spectra at NOMAD,
    within a transverse fiducial area of $260 \times 260$~cm$^2$, as
    predicted by the NOMAD simulation of the neutrino beam line.}
  \label{fig:neutcomp}
\end{figure}

The resulting energy spectra of $\numu$ and $\nue$\,, and of their
components, are shown in Fig.~\ref{fig:neutcomp}.  The $\numu$ flux
is predominantly due to decays of $\pip$ up to 60 GeV neutrino energy
and to those of $\kp$ beyond this energy. The bulk of the
$\nue$ flux comes from the decays of $\kp$, with $\klong$ contributing
at the level of about 18\% and $\mup$ at the level of about 14\%.
The composition of the beam is shown in Table~\ref{tab:flav-cont}.

\begin{table}[hbt]
  \vspace*{0.2cm}
  \caption{Average energies and relative abundances of the fluxes and
    charged current events of the four principal neutrino flavours at NOMAD,
    within a transverse fiducial area of $260 \times 260$~cm$^2$.}
  \begin{center}
    \begin{tabular}{cccccc}
      & \multicolumn{2}{c}{Flux} & & \multicolumn{2}{c}{CC interactions} \\ \cline{2-3} \cline{5-6}
      Flavour & $\langle E_{\nu}\rangle~[{\rm GeV}]$ & Rel.\ abund. & &
      $\langle E \rangle~[{\rm GeV}]$ & Rel.\ abund. \\ \hline
      $\numu$  &  24.3  &  1.0     &  &  47.5  &  1.0     \\
      $\anumu$ &  17.2  &  0.068   &  &  42.0  &  0.024   \\
      $\nue$   &  36.4  &  0.010   &  &  58.2  &  0.015   \\
      $\anue$  &  27.6  &  0.0027  &  &  50.9  &  0.0015  \\ \hline
    \end{tabular}
  \end{center}
  \label{tab:flav-cont}
\end{table}

An alternative method of predicting the $\nue$ content of the beam,
% is currently nearing completion.
not used in this paper, has also been studied.
This method, referred to as the Empirical Parameterization, does not use
hadronic interaction packages such as FLUKA to predict the yield of
particles from p-Be interactions.  Instead it uses the $\numu~\cc$,
$\anumu~\cc$ and $\anue~\cc$ events observed in NOMAD as well as
the results of NA20 and NA56 to constrain the hadron production
cross sections, and using these, to predict the $\nue$ content of
the beam.
% An analysis of the NOMAD data based on this method will
% be reported in a later paper.

\section{Event simulation}\label{sec:sim}
The neutrino flux generated by NUBEAM was used as an input to the NOMAD
event generator to produce interactions of $\numu$, $\anumu$, $\nue$ and
$\anue$.  Deep-inelastic scattering (DIS) events were simulated with
a modified version of the LEPTO 6.1 event generator~\cite{LEPTO}, with
$Q^2$ and $W^2$ cutoffs removed.  Quasi-elastic (QEL)~\cite{QEL} and
resonance production (RES)~\cite{RES} events were generated as well.
The GRV-HO parameterization~\cite{GRVHO}
of the parton density functions and the nucleon Fermi motion distribution
of Ref.~\cite{PFermi}, truncated at 1~GeV/$c$, were used along with
JETSET 7.4~\cite{JETS} to treat the fragmentation.

The secondary particles produced in these interactions were then
propagated through a full GEANT3~\cite{GEANT} simulation of the detector.
The size of the simulated samples exceeded the data samples by about a factor
of three for $\numu~\cc$, 10 for $\anumu~\cc$ and neutral current (NC)
and 100 for $\nue~\cc$ interactions.

The contributions of QEL, RES and DIS events in the Monte Carlo were
adjusted to reproduce the $W^2$ distribution of $\numu~\cc$ interactions
observed in the data, taking into account the non-isoscalarity
of the NOMAD target (52.4\% protons and 47.6\% neutrons).  The
fractions of QEL and RES events included in the Monte Carlo are listed
in Table~\ref{tab:qel-res}.

\begin{table}[thb]
  \vspace*{0.2cm}
  \caption{Estimated percentages of QEL and RES events in the NOMAD data.}
  \begin{center}
    \begin{tabular}{ccccccccccc}
      \multicolumn{2}{c}{$\numu$} & & \multicolumn{2}{c}{$\anumu$} & &
      \multicolumn{2}{c}{$\nue$}  & & \multicolumn{2}{c}{$\anue$} \\
      \cline{1-2} \cline{4-5} \cline{7-8} \cline{10-11}
   QEL  &  RES  & &  QEL  &  RES  & &  QEL  &  RES  & &  QEL  &  RES  \\ \hline
  2.5\% & 3.3\% & & 6.3\% & 7.1\% & & 1.7\% & 2.2\% & & 4.3\% & 4.8\% \\ \hline
    \end{tabular}
  \end{center}
  \label{tab:qel-res}
\end{table}

\section{Data collection and analysis}\label{sec:data}
NOMAD collected data from 1995 to 1998.  Most of the running, a total
exposure of $5.1 \times 10^{19}$ protons on target (pot), was in neutrino
mode and yielded $1.3 \times 10^6$ $\numu~\cc$ interactions in the fiducial
volume of the detector.  Some data,
amounting to $0.44 \times 10^{19}$ pot, were also collected in antineutrino
mode and some, $0.04 \times 10^{19}$ pot, in zero-focusing mode (with the horn
and reflector switched off); these data were used mostly to check the
beam simulation.

The trajectories of charged particles are reconstructed from the hits
in the drift chambers and, from these trajectories, momenta are computed
using the Kalman filter technique~\cite{Kalman}, which accounts for
energy loss along the
trajectory.  As a first step the energy loss model used is that for
pions, resulting in a momentum estimate, $p_\pi$, at the beginning
of the track.  Particles later identified as electrons or positrons
are refitted~\cite{BP} with an additional average energy loss due
to bremsstrahlung,
resulting in a new estimate, $p_e$, of the momentum.  Energy clusters
in the ECAL not associated to charged particles are assumed to be
due to photons.

Vertices are reconstructed from the trajectories of charged particles.
The energy of the incident neutrino, $E_\nu$, is approximated by the
total (visible) energy of an event computed from the sum
of the energies of all observed primary particles and of photons.
The reconstruction efficiency for the hadronic jet was found to be
overestimated by the Monte Carlo.  This was due mostly to reconstruction
effects such as the density cut described in Section~\ref{sec:evsel},
as well as an inadequate
treatment of nuclear reinteractions (the interactions of produced
particles inside the nucleus in which the neutrino interaction
occurred) and a harder fragmentation in the Monte Carlo than
observed in the data.  The resulting overestimate of the scale of the hadronic
energy, $\Eh$, could be studied by noting that the differential
cross section for charged current neutrino interactions is almost independent
of $\Ybj=\Eh/E_\nu$.  This entails that the distribution of $R_E$,
the ratio of the
average neutrino energy in the data to the average neutrino energy
in the Monte Carlo, will be independent of $\Ybj$ if $\Eh$ is correctly
measured.  However, if the measured $\Eh$ is systematically reduced by
an amount $\alpha$ with respect to the true value of $\Eh$, then the
dependence of $R_E$ on $\Ybj$ can be described by a straight line with slope
($1-\alpha$) for small values of $\alpha$.  Fitting the $R_E$ distribution
as a function of $\Ybj$ allowed us to determine that the correction to
$\Eh$ is $\alpha = (8.3 \pm 1.5)$\%.

Since the electron radiates bremsstrahlung photons in traversing the
drift chambers, in order to have an accurate measure of its energy,
these photons must be identified and their energy added to the energy
of the ECAL cluster at the end of the electron trajectory.  Because
of the curvature of the electron trajectory in the magnetic field
these photons are located in a vertical fan delimited, on the one hand,
by the actual trajectory of the electron between the event vertex and
the point of impact of the electron on the ECAL and, on the other hand, by
the extrapolation of the initial direction of the electron to the ECAL.
The energy of photons in the ECAL and of photon conversions in the DC
found in this region is included, resulting in a measure of the electron
energy, $\Ebrem$, with an average resolution of 2.1\%.

\section{Principles of oscillation search}\label{sec:principles}
The $\nmne$ oscillation signal should manifest itself as an excess in the
number of $\nue~\cc$ events over that expected for an intrinsic $\nue$
contamination in the beam (about 1\% of $\numu$).   In order
to reduce systematic uncertainties associated with absolute flux
predictions and selection efficiencies, we study the ratio $\Rem$
of the number of $\nue$ to $\numu$ charged current interactions.
Due to different energy and radial distributions of incident electron
and muon neutrinos, the contribution of the intrinsic $\nue$ component
is smaller at low $\nue$ energies, $E_\nu$, where a low $\dm$ oscillation
signal is expected, and at small radial distances from the beam axis, $r$.
Thus, the sensitivity of the search is increased by taking into account the
dependence of $\Rem$ on $E_\nu$ and $r$.

The presence or absence of $\nmne$ oscillations is established by comparing
the measured $\Rem$ with the one expected in the absence of oscillations.
In order to avoid biases, we adopted a ``blind analysis'' strategy:
the comparison of the measured to the predicted $\Rem$ is not made
until the accuracy of the flux predictions and the robustness of the
data analysis have been demonstrated and until all selection criteria
are fixed.  A number of control data samples in which no oscillation
contribution is expected (charged current interactions of $\numu$,
$\anumu$ and $\anue$ in neutrino mode, and  of $\numu$, $\anumu$,
$\anue$ and $\nue$ in antineutrino and zero-focusing modes) are used
to verify the flux predictions and the validity of the Monte Carlo
simulation~\cite{fluxes}.  It should be noted that no oscillation
signal is expected to be measurable in $\anue$ since the intrinsic ratio
of $\anumu/\anue$ in the beam is four times smaller than the intrinsic
$\numu/\nue$ ratio and the antineutrino statistics is limited.
% and the rate of $\anue~\cc$ interactions is ten times
% smaller than that of $\nue$. 

\section{Event selection}
\label{sec:evsel}
In order to calculate $\Rem$, pure samples of $\nue~\cc$ and $\numu~\cc$
interactions are selected.
The initial data sample for $\nue~\cc$ interactions is complementary
to that used in the $\numu~\cc$ selection described below, i.e.,
it consists only of those
events that include no muon (identified with looser criteria than in the
$\numu~\cc$ selection).  The basic requirement is the presence of a track
associated with the neutrino interaction vertex, pointing to an energy
deposition in the ECAL and identified as an electron in the TRD and ECAL.
The identification criteria are:
\begin{itemize}
\item Pulse heights in the TRD consistent with those of an electron
  and such that isolated charged pions are rejected by at least a
  factor of 1000.
\item A momentum-energy match satisfying:
  \begin{itemize}
  \item $(\Ebrem-p_{\pi})/(\Ebrem+p_{\pi})\;>\;-0.3$;
  \item $(\Ebrem-p_{e})/(\Ebrem+p_{e})\;<\;0.4$;
  \item no activity in HCAL associated with the electron trajectory.
  \end{itemize}
\end{itemize}

\begin{figure}
  \vspace*{-0.5cm}
  \begin{center}
    \includegraphics*[width=1.\textwidth]{\epsdir/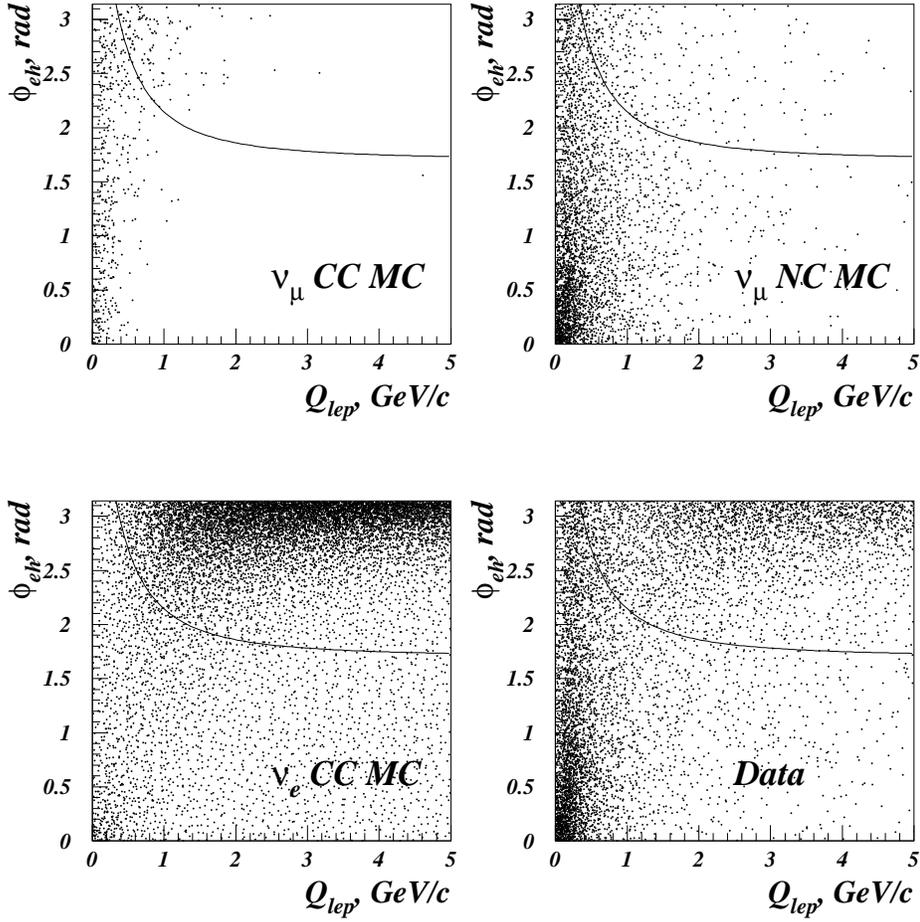}
  \end{center}
  \vspace*{-0.3cm}
  \caption{The two-dimensional distributions $\phieh$-$\qlep$
    (defined in the text) for Monte Carlo $\numu~\cc$ and $\numu~\nc$
    (background) events and for $\nue~\cc$ (signal) events, as well as
    for the NOMAD data.  The events to the right of the curve were
    selected.}\label{fig:phi_qlep}
\end{figure}

Electrons from conversions and Dalitz decays are rejected by requiring:
\begin{itemize}
\item that the first point on the candidate track be within 15~cm
  of the primary vertex;
\item that no positively charged track, either identified as a positron
  in the TRD or missing the TRD, when taken
  together with the candidate electron, results in the combination
  being consistent with a conversion. The criteria used are the
  invariant mass and the angle between the plane containing the
  trajectories of the two tracks and the vertical.
% coplanarity of the two tracks with the vertical plane.
\end{itemize}
In order to reduce further the background from neutral current and
charged current events in which the muon was not identified, kinematic
cuts are also applied using the following two variables:
\begin{itemize}
\item $\phieh$, the angle between the electron and the hadronic jet
  in the plane transverse to the neutrino beam direction.
\item $\qlep$, the component of the electron momentum perpendicular
  to the had\-ronic jet direction.
\end{itemize}
Charged pions simulating secondary electrons and conversion electrons
are part of the hadronic jet, resulting in small values of these variables.
Primary electrons are isolated resulting in large values of $\phieh$ and
$\qlep$.  These differences are evident in the $\phieh$-$\qlep$ plots
shown in Fig.~\ref{fig:phi_qlep} on which the cut used is also shown.

Single track events, originating mostly from quasi-elastic and resonance
production interactions, are treated somewhat differently because they are
more prone to background arising from charged particles entering the
detector without being registered in the veto counter and because no
hadronic jet containing primary charged particles is present.  The criteria
used to select them are:
\begin{itemize}
\item no activity in the drift chambers upstream of the beginning
  of the track and consistent with having given rise to the track;
\item the angle between the single electron and neutrino beam directions,
  $\theta_{\nu e}$, smaller than 150~mrad;
\item $\Ybj < 0.5$;
\item $p_e \times \theta_{\nu e}^{2} > 2m_e$, with $m_e$ the mass of
  the electron, a cut that rejects neutrino-electron scattering events.
\end{itemize}

Lastly, only events with $p_e > 2.5$~GeV/$c$ and $E_\nu < 300$~GeV are
retained.  These selection criteria result in an efficiency for
$\nue~\cc$, estimated from the Monte Carlo, of 43.9\%.
% Averaged over all data-taking periods using the numbers of pots.

The surviving background contribution to the $\nue~\cc$ sample is estimated,
from the Monte Carlo, to be 1.8\%.  It consists mostly of electrons
from photon conversions.  Their rejection depends critically on the
reconstruction of the accompanying positron and on identifying a conversion
point distinct from the primary vertex.  The reconstruction of very low
energy positrons and the separation of a conversion vertex from the primary
vertex in a high multiplicity event could be different in the data and in
the Monte Carlo.  As a cross-check, a class of $e^+$ events that
failed the kinematic cuts or were produced far from the primary vertex,
and thus consisting almost entirely of background, were selected in
both the data and in the Monte Carlo.  The number of such events was
found to be higher in the data by a factor that varied between 1.0
and 1.6 depending on $E_\nu$.  The Monte Carlo background estimate
for the $e^-$ events was therefore multiplied by this same factor,
thus raising the total background estimate to 2.3\%.

Charged current interactions of $\numu$ are characterized by the presence
of a primary muon in the final state, which had to penetrate 13 interaction
lengths of absorber material to reach both muon
stations in order to be identified.
In addition, in order to minimize the differences between selection
efficiencies of $\numu~\cc$ and $\nue~\cc$ events, we apply kinematic
criteria identical to those used in the $\nue~\cc$ selection, although
they are not needed for the background suppression.  The resulting
$\numu~\cc$ data sample has a negligible
background contamination; the average selection efficiency is 60\%. % 59.8\%

\begin{figure}
  \begin{center}
    \begin{minipage}[c]{0.5\textwidth}
      \centering\includegraphics[width=1.\textwidth]{\epsdir/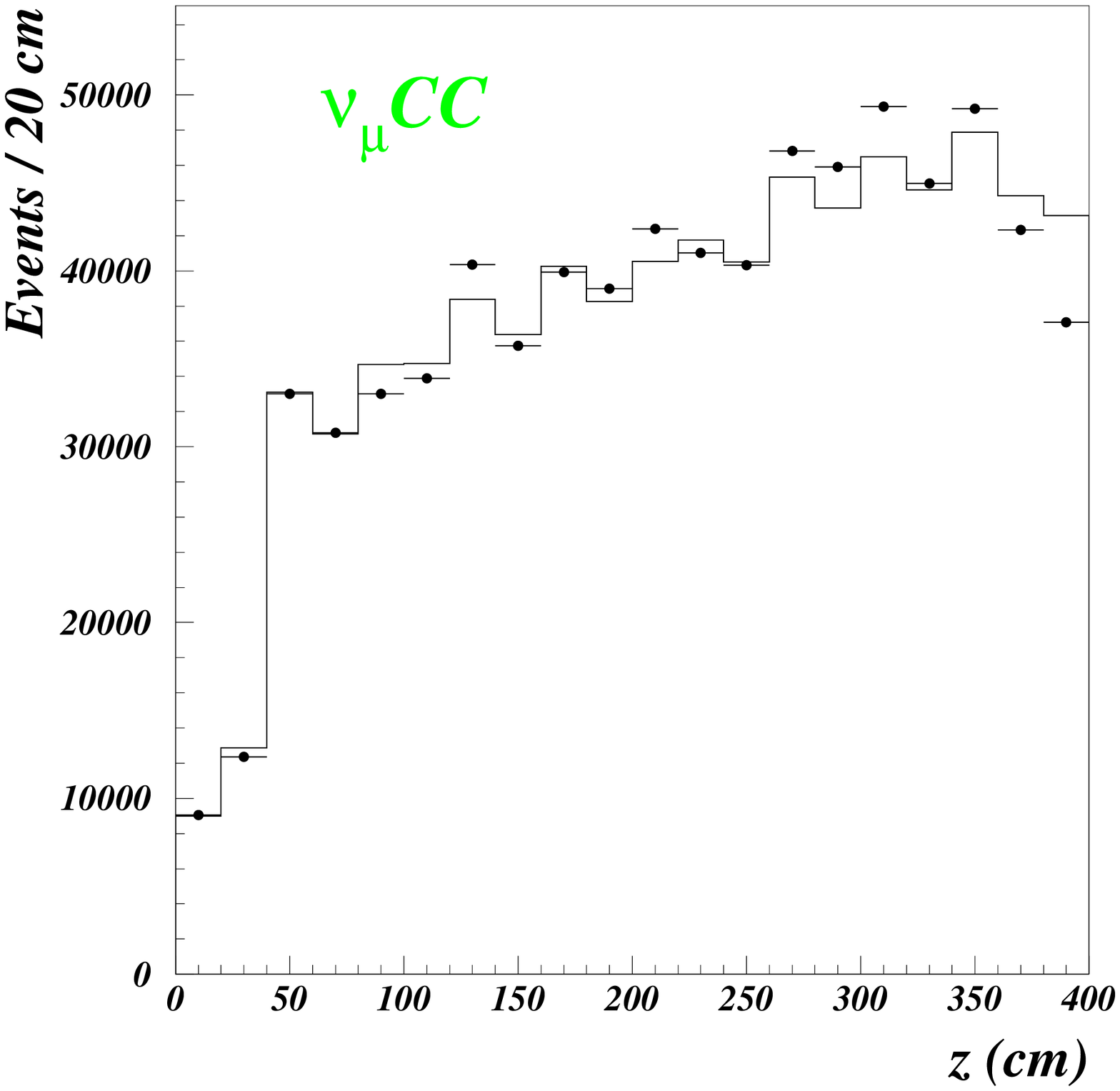}
    \end{minipage}\hfill
    \begin{minipage}[c]{0.5\textwidth}
      \centering\includegraphics[width=1.\textwidth]{\epsdir/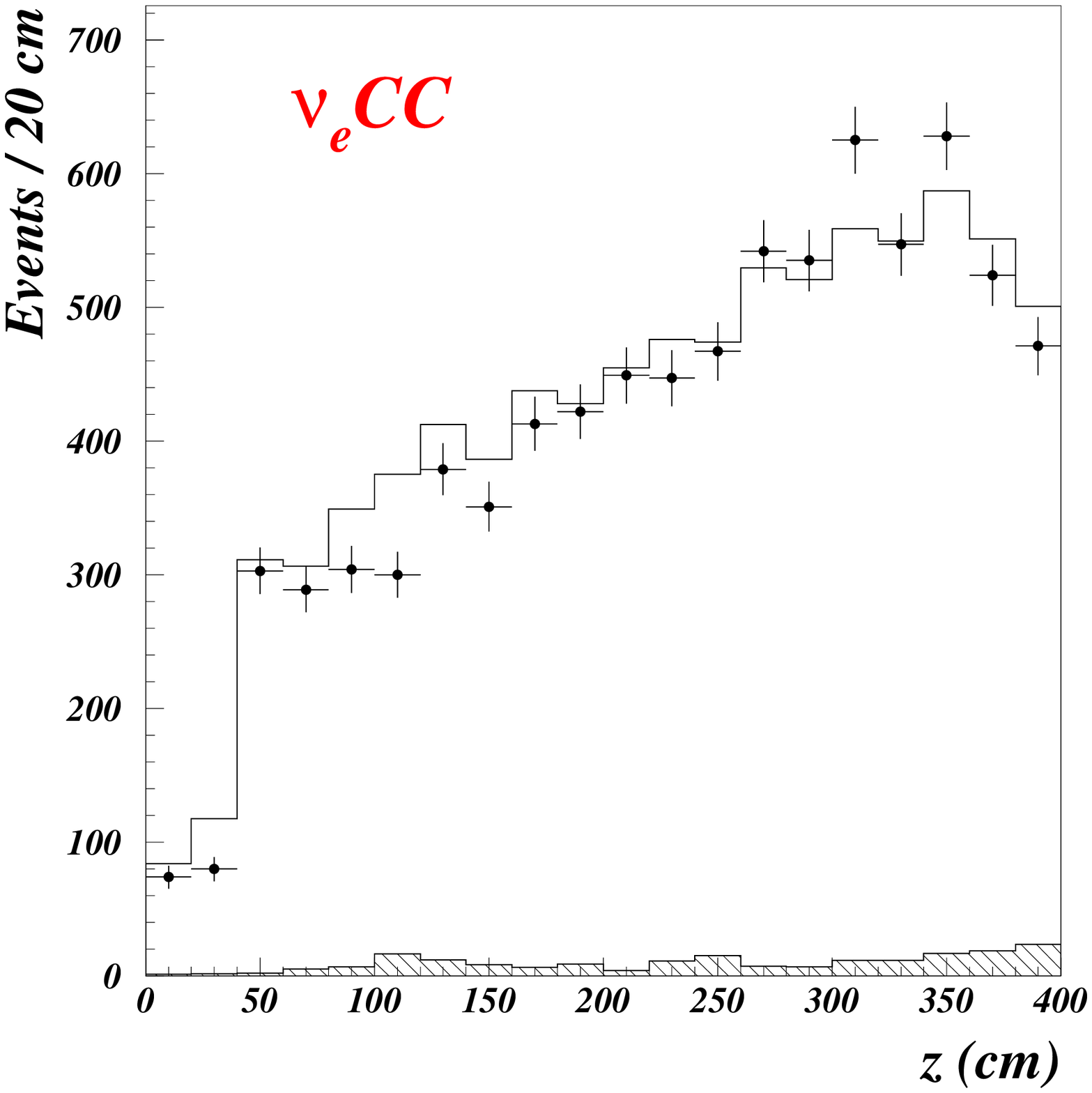}
    \end{minipage}
  \end{center}
  \vspace*{0.2cm}
  \caption{The distributions of $z$ (defined in the text) for $\numu~\cc$
    (left) and $\nue~\cc$ (right) candidates in the data (points with error
    bars) and in the Monte Carlo (histogram).  The Monte Carlo distribution
    of $\numu~\cc$ events is normalized to the number of $\numu~\cc$ events
    in the data; the Monte Carlo distribution of $\nue~\cc$ events is
    normalized using the relative $\nue~\cc/\numu~\cc$ abundance predicted
    by our simulation.  The background contribution
    to the $\nue~\cc$ sample is shown in the hatched histogram.}
  \label{fig:cc_zvtx}
\end{figure}

The geometrical and kinematical distributions of both types of events
are well reproduced by the Monte Carlo simulation, with the exception of the
distribution of interaction vertices along $z$, the beam direction. The data
and Monte Carlo $z$ distributions are shown in Fig.~\ref{fig:cc_zvtx}
for $\numu~\cc$ and $\nue~\cc$ events. It can be seen that fewer
events are present in the data than in the Monte Carlo at small $z$
especially for $\nue~\cc$ events.  The origin of this difference is
due mostly to a cut introduced during the reconstruction
of events: events with a very high density of hits in the drift
chambers were not reconstructed due to a prohibitive reconstruction
time.  Since the data has on average a higher density of hits than the
Monte Carlo, the effect of this cut is different on the two
samples.  Furthermore, since electrons radiate photons in traversing
the drift chambers and some of these photons convert, the
density of hits in $\nue~\cc$ events is large thus exacerbating the
effect of the density cut for these events.  The reprocessing of a sample
of data and Monte Carlo events without this density cut resulted in $z$
distributions that were in much better agreement.  We therefore decided
to restrict the analysis to events occurring in the 72 downstream planes
of drift chambers by requiring $z > 184$ cm.  This restriction
resulted in a loss of about 30\% of the $\nue~\cc$ events and 35\% of
the $\numu~\cc$ events.  It should
be noted that any oscillation effects could not manifest themselves
over this distance since the point of origin of the neutrinos is
spread over more than 300~m.

\begin{figure}
  \begin{center}
    \begin{minipage}[c]{0.5\textwidth}
      \centering\includegraphics[width=1.\textwidth]{\epsdir/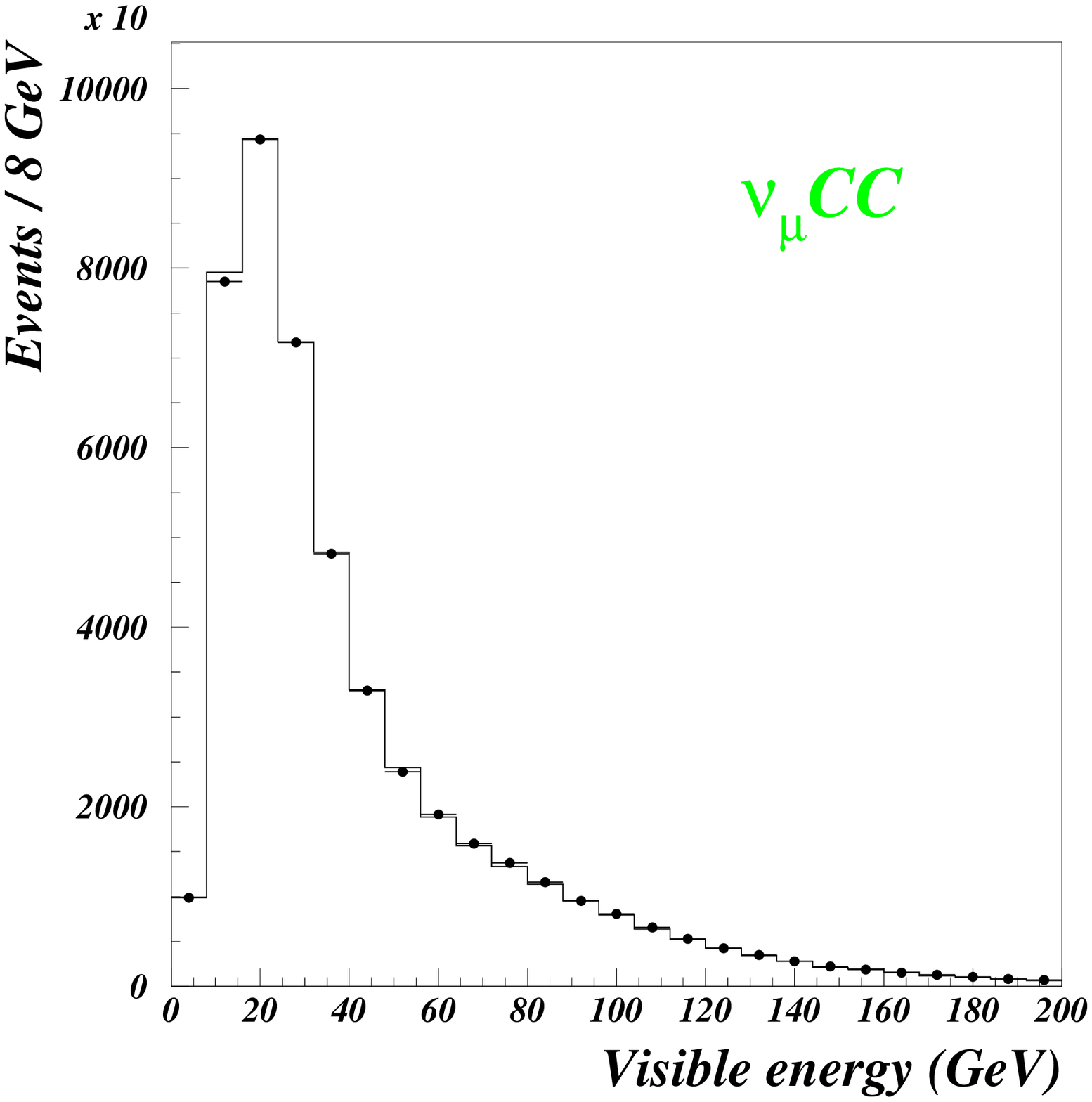}
    \end{minipage}\hfill
    \begin{minipage}[c]{0.5\textwidth}
      \centering\includegraphics[width=1.\textwidth]{\epsdir/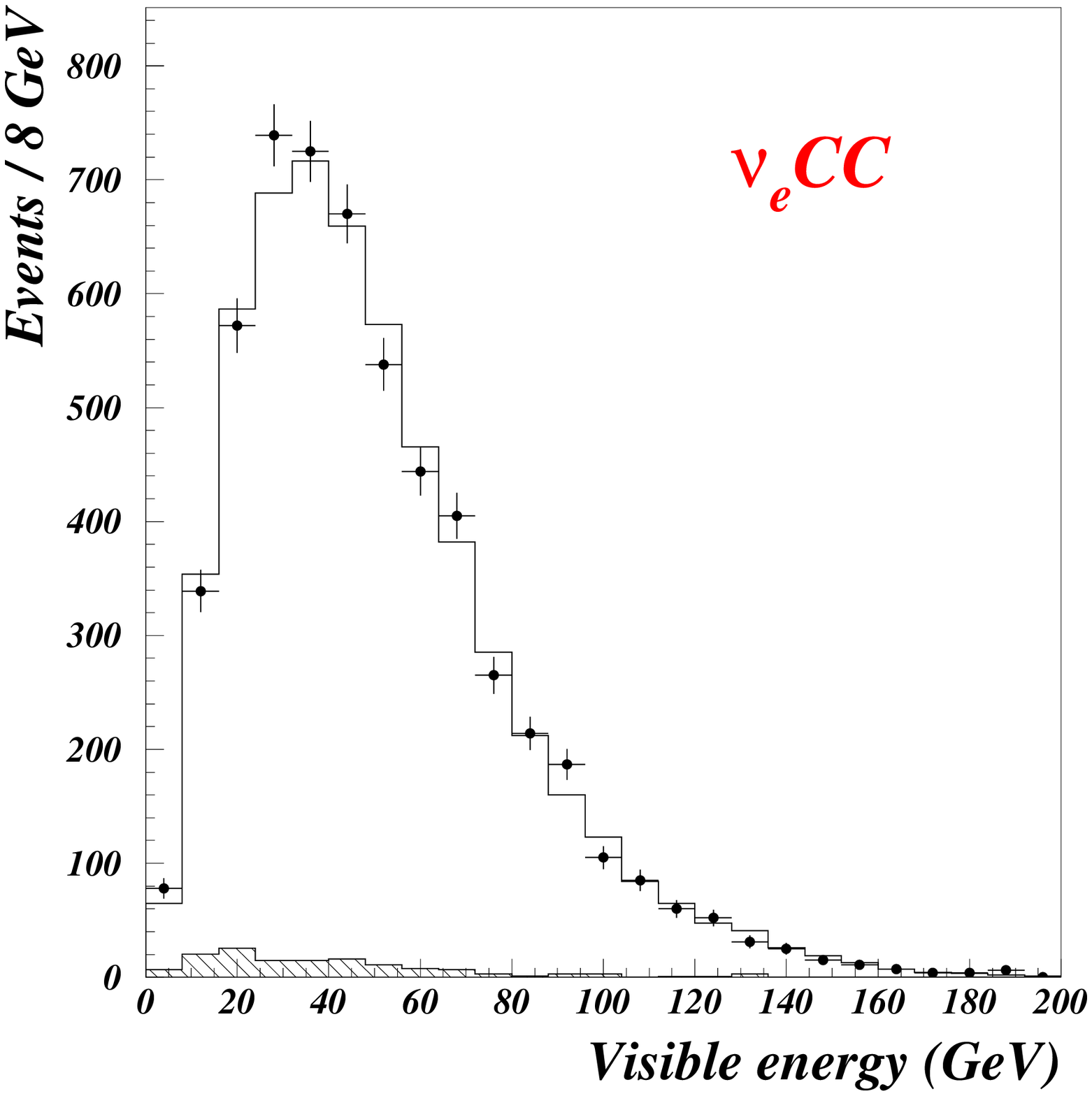}
    \end{minipage}
  \end{center}
  \vspace*{0.2cm}
  \caption{Neutrino energy spectra for the data (points with error bars)
    and the Monte Carlo (histogram), for $\numu~\cc$ (left) and $\nue~\cc$
    (right) candidates.  The normalization of the Monte Carlo distributions
    is the same as in Fig.~\protect\ref{fig:cc_zvtx}.  The background
    contribution to the $\nue~\cc$
    sample is shown in the hatched histogram.}\label{fig:cc_evis}
\end{figure}

\begin{figure}
  \begin{center}
    \begin{minipage}[c]{0.5\textwidth}
      \centering\includegraphics[width=1.\textwidth]{\epsdir/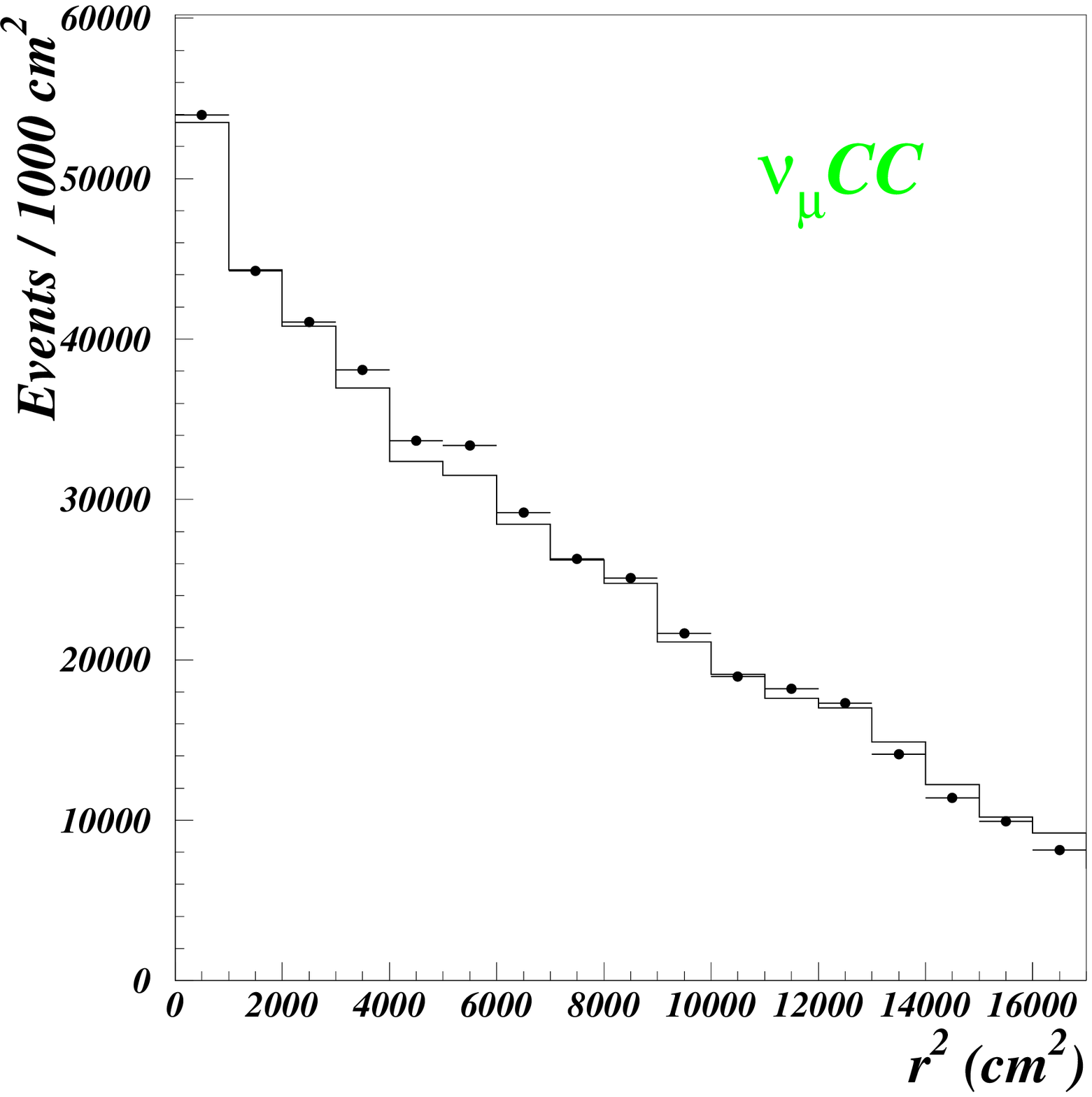}
    \end{minipage}\hfill
    \begin{minipage}[c]{0.5\textwidth}
      \centering\includegraphics[width=1.\textwidth]{\epsdir/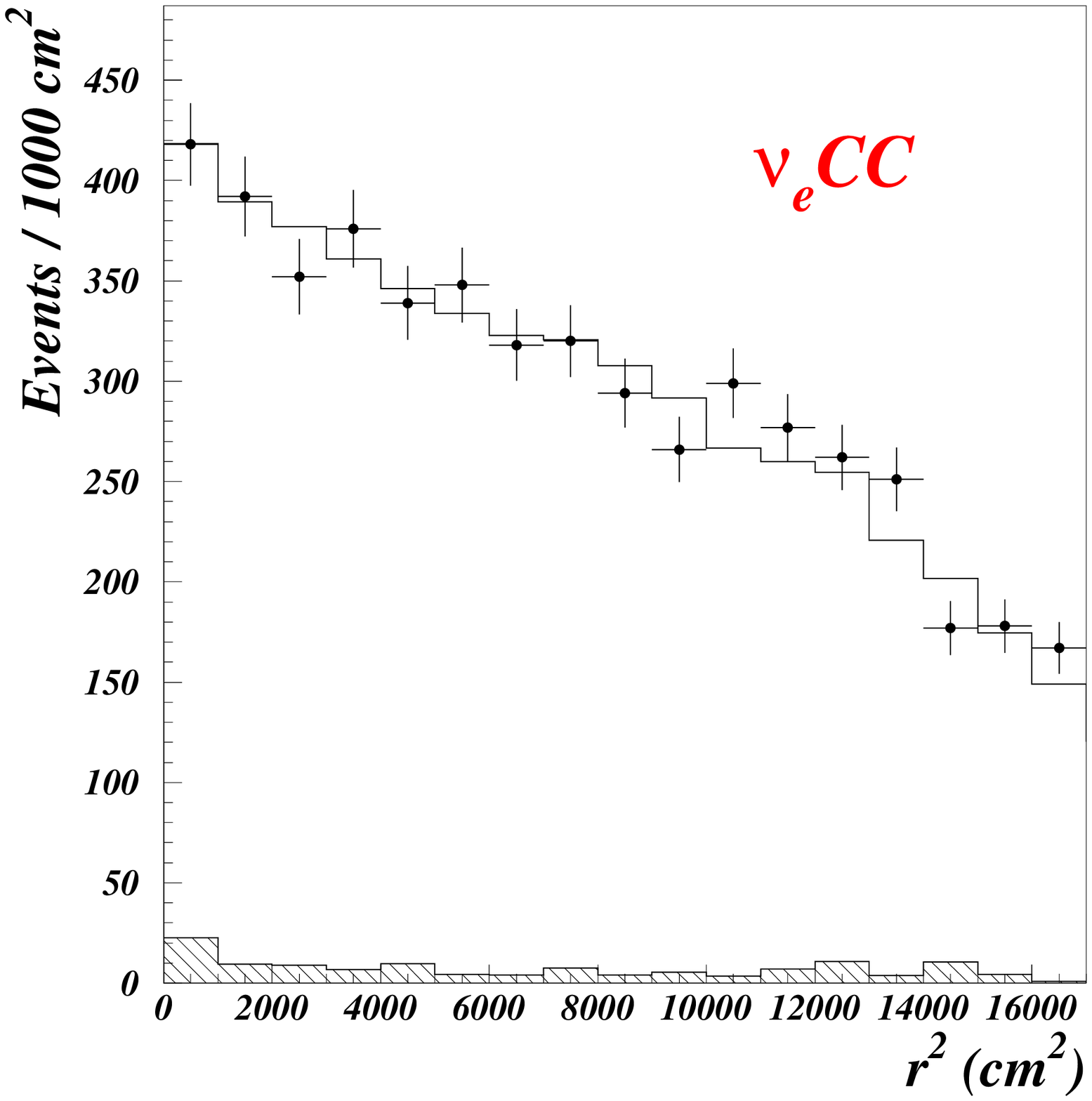}
    \end{minipage}
  \end{center}
  \vspace*{0.2cm}
  \caption{The distribution of $r^2$, the square of the radial position
    of the neutrino interaction vertex with respect to the nominal beam axis,
    for the data (points with error bars) and the Monte Carlo (histogram),
    for $\numu~\cc$ (left) and $\nue~\cc$ (right) candidates. The normalization
    of the Monte Carlo distributions is the same as in the two previous plots.
    The background contribution to the $\nue~\cc$ sample is shown in the
    hatched histogram.}\label{fig:cc_r2}
\end{figure}

A total of 5,584 $\nue~\cc$ and 472,378 $\numu~\cc$ events were retained.
Their energy spectra and radial distributions are shown in
Fig.~\ref{fig:cc_evis} and Fig.~\ref{fig:cc_r2}, respectively.

\section{Systematic uncertainties}
\begin{figure}[htb]
  \vspace*{-0.3cm}
  \begin{center}
    \includegraphics[width=1.\textwidth]{\epsdir/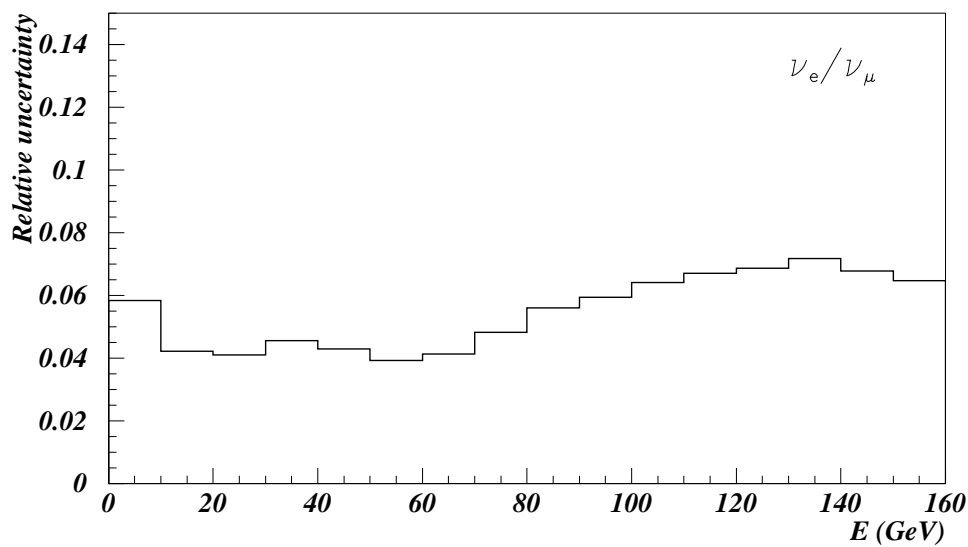}
  \end{center}
  \caption{Energy-dependent uncertainty in the prediction of the $\Rem$ ratio.}
  \label{fig:err_ratio}
\end{figure}

The single largest uncertainty in this oscillation search is the
uncertainty in the prediction of the fraction and the energy spectrum
of intrinsic $\nue$
present in the beam.  The computation of this uncertainty is described
in detail in Ref.~\cite{fluxes}.  Most of it is due to the limited
knowledge of the particle production cross sections in p-Be interactions.
In turn, this is due to the small number of experimental data points
on $\pip$ and $\kp$ production measured by NA20 and NA56, especially
at non-zero values of transverse momentum.  This uncertainty is
energy-dependent; its typical fractional value at low $E_\nu$ is 4\%.
The second largest systematic error of about 3\% comes from an % 15\%
uncertainty in the production of $\klong$ %content of the secondary beam,
which accounts for 18\% of the $\nue$ flux.  Other potential sources
of errors (such as tertiary particle yields, variations in the horn
current, misalignments of the focusing devices and collimators, or
inaccuracies in the simulation of the beam line elements) have also been
investigated~\cite{fluxes}; their cumulative contribution is
about 3\%.  The overall uncertainty arising from the knowledge
of the beam composition is divided into an energy-independent, or
normalization, uncertainty and an energy-dependent one.  The normalization
uncertainty on $\Rem$ is 4.2\%, while the energy-dependent uncertainty,
shown in Fig.~\ref{fig:err_ratio}, varies from 4\% to 7\%.

The following additional sources of systematic uncertainties arising from the
data selection and analysis were studied:
\begin{itemize}
\item The correction to the background contribution. By studying the
  $e^+$ events used to determine this correction, the uncertainty
  attached to it was found to be 15\%, resulting in an uncertainty
  of less than 1\% in $\Rem$.
\item The electron reconstruction and identification.  The electron
  reconstruction efficiency in the drift chambers ($\sim 98$\%) was
  studied by defining electron tracks using the TRD, preshower and
  ECAL only, and then computing the frequency that a DC track is associated
  to the electron track.  The electron identification efficiency of the TRD
  ($\sim 93$\%) was studied using $\delta$-rays produced by isolated muons
  arising from a nearby test beam and traversing the NOMAD detector outside
  of the neutrino spill.  The efficiency of the momentum-energy consistency
  check % ($\sim 98$\%) - this is only for the first cut of the two
  was studied using $e^{\pm}$ from photon
  conversions.  Taken together, the uncertainties in the reconstruction
  and identification efficiencies for electrons result in a 1\%
  uncertainty in $\Rem$.
\item A possible relative shift between the electron and muon energy
  scales.  The upper limit on the absolute energy uncertainty of the ECAL was
  0.5\%~\cite{ECAL_nonlin}, resulting in an uncertainty of less than 1\% in
  $\Rem$.  Non-linearity corrections to the ECAL energy scale have negligible
  effects on $\Rem$.  The muon momentum scale introduced a negligible error.
\item The mixture of QEL, RES and DIS events in the Monte Carlo.
  Cross sections for QEL and RES processes were varied by as much
  as 50\% resulting in a negligible effect on $\Rem$.
\item The fragmentation model.  The effect on $\Rem$ of the fragmentation
  and nuclear reinteraction model used in the Monte Carlo was studied by
  comparing the prediction for $\Rem$ of two such models.  Differences of
 % the order of a few \%
  up to 2\% were observed and introduced as an energy-dependent
  systematic uncertainty.
\item The kinematic (isolation) cuts.  This was studied by tightening
  and loosening the cuts in both the data and
  Monte Carlo and studying the effect on $\Rem$.  Since this was
  a blind analysis, this study could only be made after the analysis was
  frozen and the electron neutrino data examined.  The effect was
  negligible in the relevant neutrino energy range, from 10 to 80~GeV.
\item The hadronic energy scale.  The 18\% uncertainty on $\alpha$, the
  hadronic energy correction, resulted in less than a 1\% uncertainty
  on $\Rem$ throughout the neutrino energy range.  It was included as
  an energy-dependent uncertainty.  The uncertainty arising from the
  method of calculating the hadronic energy was studied using
  two alternative methods to calculate it -- the Myatt
  method~\cite{Myatt} and the double-angle method~\cite{Bentvelsen}, which use
  both the energy and direction of the lepton and only the direction of the
  hadronic jet.  After computing the appropriate energy correction for each
  of these methods, no systematic differences were observed between these
  alternative methods and the method described in Section~\ref{sec:data}.
  Again, this study could only be performed after examining the electron data.
\end{itemize}

\section{Results}
\begin{figure}[htb]
  \centering
  \vspace*{-0.3cm}
  \begin{minipage}[c]{0.6\textwidth}
    \centering\includegraphics[width=9.5cm]{\epsdir/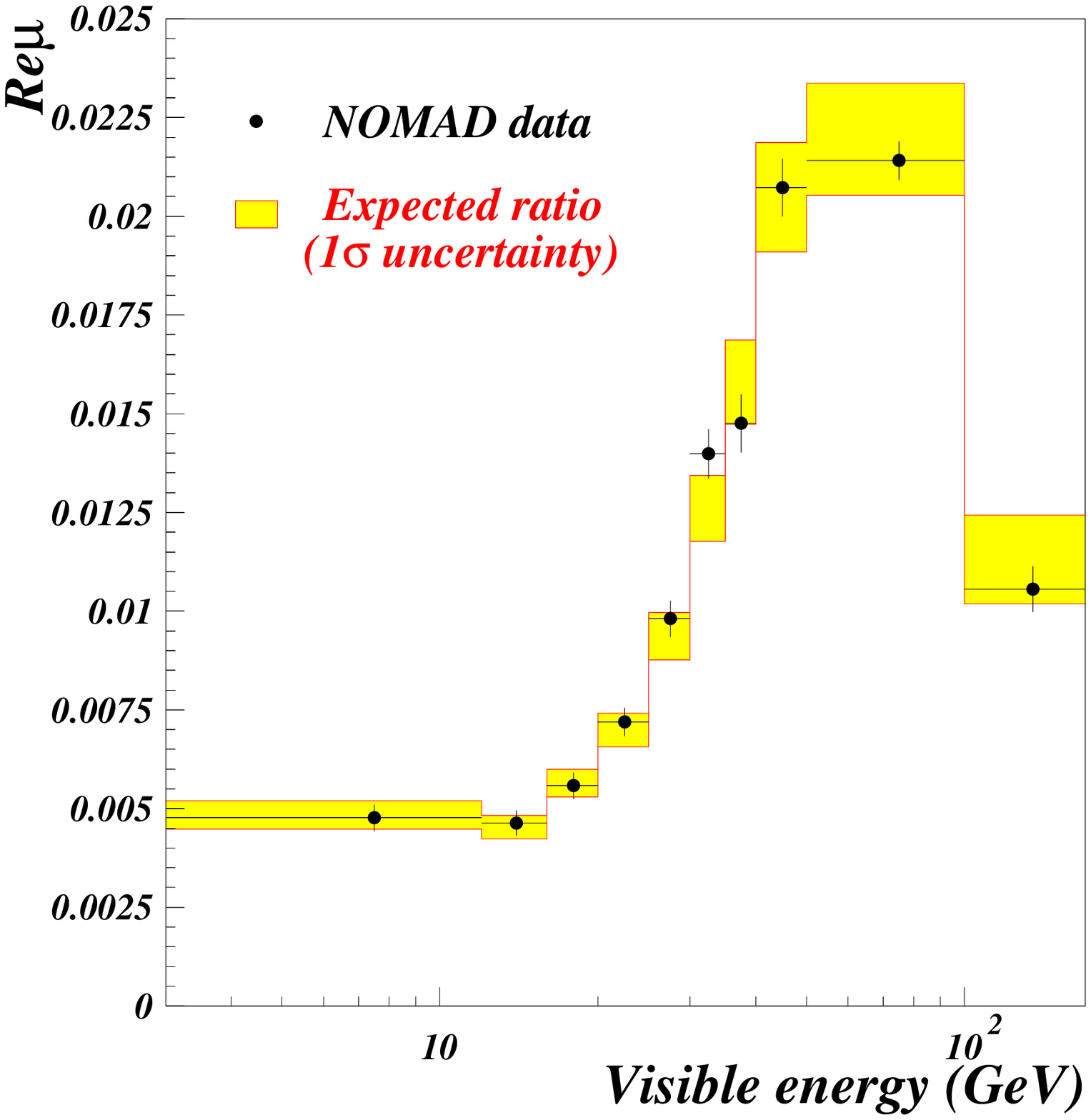}
  \end{minipage}%
  \begin{minipage}[c]{0.4\textwidth}
    \centering\includegraphics[width=5cm]{\epsdir/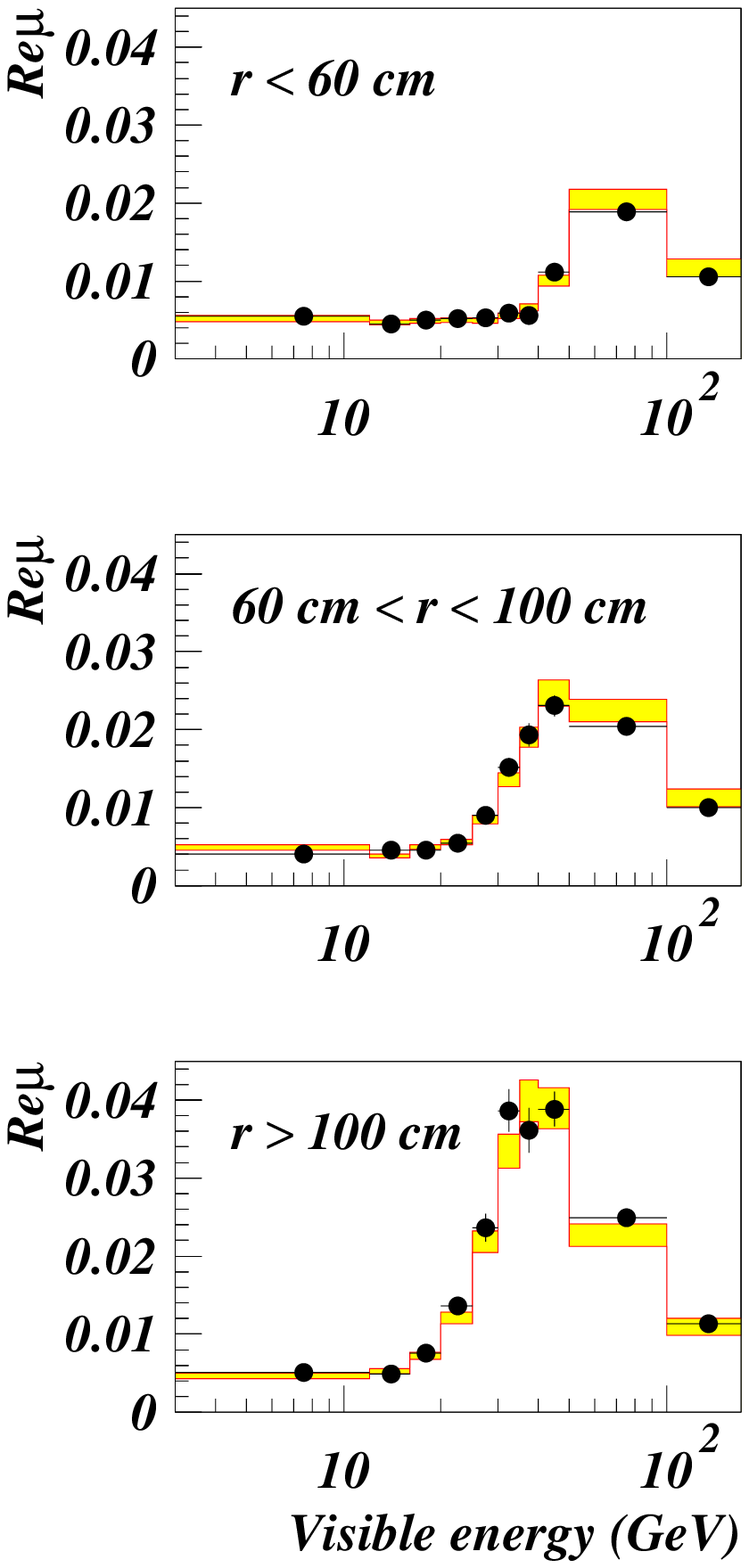}
  \end{minipage}
  \vspace*{0.1cm}
  \caption{The $\Rem$ ratio as a function of the visible energy for the
    data (points) and for the Monte Carlo prediction assuming no oscillations
    (filled bands), for the full radial acceptance (left) and in the three
    radial bins (right).  The upper and lower boundaries of the bands
    correspond
    to the predictions with $\pm 1\sigma$ uncertainty, where $\sigma$ includes
    both the normalization and energy-dependent systematic uncertainties
    added in quadrature.}
  \label{fig:ratio_data_mc}
\end{figure}

The $\Rem$ distribution as a function of the visible energy obtained
from the data is shown in Fig.~\ref{fig:ratio_data_mc}, for the full radial
acceptance (left) and in three radial bins (right).  It is in good
agreement with the Monte Carlo prediction under the no-oscillation
hypothesis, also shown in the figure as $\pm 1\sigma$ uncertainty bands:
a $\chi^2$ of 37.1/30 d.o.f. is obtained when the data are analyzed
and compared to the simulation in the 10 energy bins and the 3 radial bins
shown in Fig.~\ref{fig:ratio_data_mc} (incorporating both statistical
and systematic uncertainties).
The best fit to $\nmne$ oscillations, in the two-family approximation,
gives a similar chi-squared value, $\chi^2_{\rm min} = 37.0/28$~d.o.f.
% The $\Rem$ distributions in the three radial bins are also shown in
% Fig.~\ref{fig:ratio_data_mc} as a function of the visible energy.

\begin{figure}
  \vspace*{-0.5cm}
  \begin{center}
    \includegraphics[width=1.\textwidth]{\epsdir/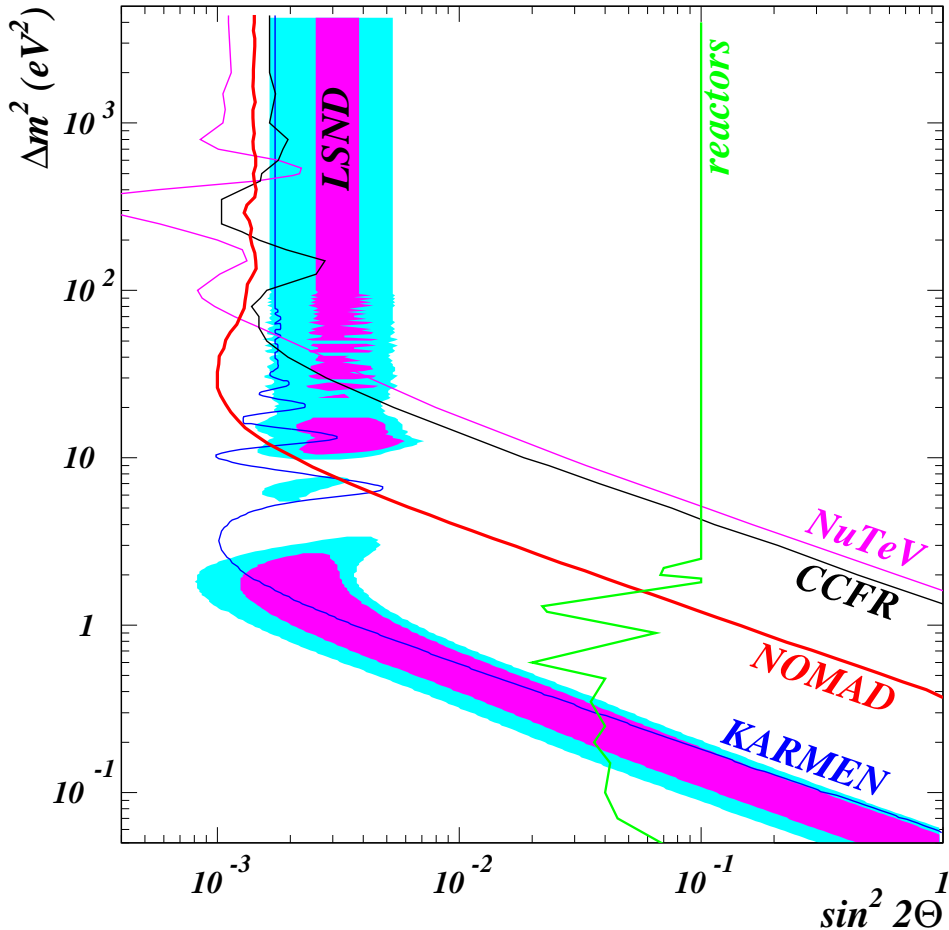}
  \end{center}
  \vspace*{-0.2cm}
  \caption{The 90\% C.L. exclusion region in the $\dm$ -- $\sintot$ plane
%    (solid line) and the sensitivity~\cite{FC} (dashed line) of this analysis,
    from this analysis superimposed on the results of other experiments.}
  \label{fig:results}
\end{figure}

We use a frequentist approach~\cite{FC} to set a 90\% confidence
upper limit on the oscillation parameters.  The resulting exclusion
region is shown in Fig.~\ref{fig:results}, together with results
of other accelerator experiments, LSND~\cite{LSND}, KARMEN~\cite{karmres},
CCFR~\cite{ccfrres} and NuTeV~\cite{nutev}, and the combined limit of
Bugey~\cite{bugey} and Chooz \cite{chooz} reactor experiments.
Values of $\dm > 0.4$~eV$^2$ for maximal mixing and
$\sintot > 1.4 \times 10^{-3}$ for large $\dm$ are excluded.  For comparison,
the sensitivity~\cite{FC} of the experiment is found to be $\dm > 0.4$~eV$^2$
for maximal mixing and $\sintot > 1.3 \times 10^{-3}$ at large $\dm$.
Our result rules out the interpretation of the LSND
measurements in terms of $\nmne$ oscillations with $\dm \gtrsim~10$~eV$^2$.

This result
is less stringent than our preliminary result~\cite{EPS01} because of a
better understanding of the systematic uncertainties arising from the
knowledge of the beam composition.  In particular, our previous analysis
had underestimated the uncertainty arising from the $\klong$ contribution
to the $\nue$ spectrum.  In addition a different split between normalization
and energy-dependent errors was implemented for the uncertainty arising
from mesons produced in secondary interactions. % We would like to stress that,
After opening the box defined for the purpose of the blind analysis,
no modifications were made either to the central value of the beam prediction
or, other than the $z > 184$~cm cut described above, to the event selection
procedure. % Furthermore,
We have ensured that including the
events at $z < 184$~cm in the analysis did not alter our limit significantly.

\section{Conclusion}
The results of a search for $\nmne$ neutrino oscillations in the NOMAD
experiment at CERN have been presented.  The experiment looked for the
appearance of $\nue$ in a predominantly $\numu$ wide-band neutrino
beam at the CERN SPS.  No evidence for oscillations was found.  The
90\% confidence limits obtained are $\dm <  0.4$~eV$^2$ for maximal
mixing and $\sintot < 1.4 \times 10^{-3}$ for large $\dm$.  This result
excludes the high $\dm$ region of oscillation parameters favoured by
the LSND experiment.

\begin{ack}
We dedicate this paper to the memory of our friend and colleague Gianni
Conforto.

The following funding agencies have contributed to this experiment:
Australian Research Council (ARC) and Department of Industry, Science, and
Resources (DISR), Australia;
Institut National de Physique Nucl\'eaire et Phy\-sique des Particules (IN2P3),
Commissariat \`a l'Energie Atomique (CEA), Mi\-nist\`ere de l'Education
Nationale, de l'Enseignement Sup\'erieur et de la Re\-cherche, France;
Bundesministerium f\"ur Bildung und Forschung (BMBF),
%(BMBF, contract 05 6DO52),
Germany;
Istituto Nazionale di Fisica Nucleare (INFN), Italy;
Institute for Nuclear Research of the Russian Academy of Sciences, Russia;
Fonds National Suisse de la Recherche Scientifique, Switzerland;
Department of Energy, National Science Foundation, %(grant PHY-9526278),
the Sloan and the Cottrell Foundations, USA.

We thank the management and staff of CERN and of all
participating institutes for their vigorous support of the experiment.
Particular thanks are due to the CERN accelerator and beam-line staff
for the magnificent performance of the neutrino beam.  We are especially
grateful to V.~Falaleev, J.-M.~Maugain and S.~Rangod for
their invaluable contribution to the design and operation of the WANF
and for their help in the simulation of the WANF beam line.  We acknowledge
the help of A.~Ferrari for the implementation of FLUKA 2000 in our beam
simulation.  We also thank our secretarial staff, J.~Barney, K.~Cross,
J.~Hebb, M.-A.~Huber, N.~Marzo, J.~Morton, R.~Phillips and M.~Richtering,
and the following people who have worked with the
collaboration on the preparation and the data
collection stages of NOMAD:
M.~Anfreville, M.~Authier, G.~Barichello, A.~Beer, V.~Bonaiti, A.~Castera,
O.~Clou\'e, C.~D\'etraz, L.~Dumps, C.~Engster,
G.~Gallay, W.~Huta, E.~Lessmann,
J.~Mulon, J.P.~Pass\'e\-ri\-eux, P.~Petit\-pas, J.~Poin\-signon,
C.~Sob\-czyn\-ski, S.~Sou\-li\'e, L.~Vi\-sen\-tin, P.~Wicht.
Finally we acknowledge the fruitful collaboration with our
colleagues from CHORUS during the setting-up, monitoring and understanding
of the beam line.
\end{ack}

% If someone wants the boldface back, he just needs to redefine \bff to be \bf
\newcommand\bff{}

\end{document}